# Further Comments on the replies given by Yoshihiko Yokoyama and Akihisa Inoue (Mater. Trans. **50** (2009) 2504-2506.) for Tsuyoshi Kajitani (Mater. Trans. **50** (2009) 2502-2503.)

Tsuyoshi Kajitani

Department of Applied Physics, Graduate School of Engineering, Tohoku University, Sendai 980-8579, Japan

Abstract

Yokoyama and Inoue gave their replies to eight questions raised by the present author with respect to their paper on "Production of $Zr_{55}Cu_{30}Ni_5Al_{10}$ glassy alloy rod of 30mm in diameter by a cap-cast technique" in 2007. Yokoyama and Inoue stressed in their reply that there were no inadequate descriptions in their paper, except for the catalog name of a TEM machine, which was a typographical error. It is still unclear why a cap-cast technique is needed to produce glassy alloy rods up to 30 mm in diameter. Lack of essential information regarding the cap-cast technique leads us no valuable assessment of the new technique. Further questions arose.

1. **Summary of the eight questions and the replies given by Yokoyama and Inoue**

   (1) Question: Insufficient evidence is given for the formation of the metallic glass phase in the arc-melted alloy buttons.

   Answer: Yokoyama and Inoue argued their own opinion by giving the DSC curve of the vitrified region of the $Zr_{55}Cu_{30}Ni_5Al_{10}$ master alloy shown in Fig.4 in their 2007 paper[2].

   (2) Question: Since the topside of the arc-melted buttons was slowly cooled and easily formed the glassy phase compared to the bottom side of the button, which was in direct contact with the copper hearth, the crystalline phase is quite likely realized there. For this reason, slow cooling is needed to stabilize the glass phase. On the other hand, quick cooling by the use of the "copper cap" operated from the top surface of the arc-melted cast rod was a key for the successful production of alloy glassy rod. There appears to be a contradiction.

Answer: No clear answer was given.

(3) Question: In order to produce a glassy alloy rod φ 30mm and 30 mm in length (volume: 21.2 cm$^3$ = 21.2 x 10$^{-6}$ m$^3$)by the cap-cast technique, 144g of alloy ingot should be thoroughly melted by arc-melting and quickly cast into a copper mold. However, it would be difficult to thoroughly melt such a large sized alloy ingot by arc-melting. Further information should be provided on the melting method.

Answer: Model: The NEV-IHC200 model made by Nisshin Giken Inc. was used.

(4) Question: Arc-melting is known to be very useful for making small sized metal ingots, typically 25g. In other words, it is very difficult to make a large sized alloy ingot that is 144g or larger in weight by the use of a typical laboratory arc-melting machine. Achieving a homogenous molten state for the successful casting was essential but that could not be realized.

Answer: Alloy buttons should weigh less than 30 g in order to ensure the best homogeneity during the arc melting. To ensure complete melting before casting, they carried out arc heating in two steps: (1) melting all the master alloy buttons together, and (2) further melting of the molten alloy just at the pouring gate, as has been described in our earlier papers (Refs.7 and 8; cited in their report[2]).

(5) Question: It is not clear why Fig.4(b) is a kind of patch-work photograph.

Answer: Yokoyama and Inoue did not have a stereoscope for observing samples with a diameter of 30mm at the time of publication. Therefore, a composite photograph with four segmented images was used to reconstruct the complete image.

(6) Question: Figure 5 shows an HREM photograph obtained by the instrument, JEM-4000FX. The authors should be careful with their documentation, because the name of the machine may actually have been the JEM-4000EX (not FX).

Answer: The instrument was a JEM-4000EX and it had high resolution in comparison to the JEM-4000FX.

(7) Question: The HREM photograph in Fig.5 is not understandable since the picture seems too smooth compared to the other HREM photographs of glassy alloys previously published. The authors are requested to give more detailed information, including how Yokoyama et al. prepared Fig.5.
Answer: Only comments on the PDF file-size were given.

(8) Question: The originality of the cap-cast technique seems questionable because production of $Zr_{55}Cu_{30}Ni_5Al_{10}$ glassy alloy rod with a diameter of 30 mm was reported in 1996 using the "suction casting method" with copper mold by Inoue *et al*. The argument "the critical size of $Zr_{55}Cu_{30}Ni_5Al_{10}$ glassy alloy rod was 20mm in diameter by the conventional technique" is misleading.
Answer: The suction casting method developed by the Inoue group in 1996 was successfully used to produce BMG rods 30mm in diameter; however, it had several disadvantages (difficulty in demolding, generation of cast defects, etc.).

**2. Critical comments on the replies**

The reply given by Yokoyama and Inoue to the comments of Kajitani [1] on the contents of their paper published in 2007 [2] are welcome and a correction has clearly been made with respect to the catalog name of the TEM machine from JEM-4000FX to JEM-4000EX. Yokoyama and Inoue tried to answer the seven questions. Nevertheless, the readers still have serious difficulties in understanding the article in 2007[2], mainly arising from lack of essential information. For this reason, nine further questions are given as follows.

(1) The paper[2] in question was reported by four researchers, while the reply[3] was written by only two of them. Do the other two, Enrico Mund and Ludwig Schultz, share the same opinion ? If this was the case, the reply[3] should have been submitted with four researchers' name.

(2) For question 1:

Yokoyama and Inoue show the DSC curve for only the vitrified part of the master ingot button. Since the DSC graph does not have a zero point or any unit on the ordinate, readers are unable to draw any quantitative conclusions about the metallic glassy phase. Also, the crystallization peak is not fully provided, so that there is no information about whether the master ingot is well prepared and how much the metallic glassy phase grows during the course of furnace cooling. Since the enthalpy of crystallization for the $Zr_{55}Cu_{30}Ni_5Al_{10}$ glassy alloy is reported as 4.3kJ/mol by Kato, Kawamura and Inoue [4], Yokoyama and Inoue could estimate the molar portion of the metallic glassy phase in the master alloy buttons. This is our new question "why they did not."

(3) For question 2:

The readers are still confused due to the contradictory argument, i.e., slow cooling is needed for the growth of the glassy phase, otherwise crystallization takes place as in the bottom part of buttons on one hand, but a solid copper "cap" is essential for producing a large scale glassy alloy rod on the other hand. No understandable explanation is given in their reply.

(4) For question 3:

The readers cannot understand "how to make the perfect alloy melt of 144 g ingot using a laboratory-scale arc-melting machine". The authors of the paper[2] are strongly requested to give a clear explanation in regard to this point. The reply[3] given by Yokoyama and Inoue just provides a photograph of the machine with the catalog values of melting capacity of 200g and the cooling water supply of 40 liters per minute at the maximum. Such information does not answer question[1] because the catalog melting capacity corresponds only to the summation of button ingots, being melted one by one (e.g. 30g) at a time in the machine at one charge. This capacity value completely differs from the maximum weight of alloy melt obtained at once, as presently requested.

Yokoyama and Inoue clearly give in their reply[3] that "melting all the master alloy buttons together" to ensure complete melting before casting. This means that they carried out arc-melting, for example, on five buttons, i.e. 30g x 5 = 150 g, all together. Again, the readers want to know "how they thoroughly melted such a large sized alloy ingot using a laboratory-scale arc-melting machine?".

Yokoyama and Inoue also indicate the experimental apparatus (NEV-IHC200 model, Nisshin Giken Inc.) and the readers are led to the conclusion that this apparatus may make it possible to melt a large amount of alloy when operating with a cooling water supply at the rate of 40 liters per minute.

According to the photograph of Fig.2[3], the size of the water tube appears to be a conventional one - 20mm in diameter - and thus the delivery rate of 40 liters per minute is impossible. In addition, the cooling water system in the Institute (IMR, Tohoku University) is confirmed to be a closed-cycle type. The difference between the inlet and outlet water pressure values does not exceed the level of 0.2 MPa, so that back pressure is essential to return the outlet water to the reservoir tank. A 20mm diameter pipe cannot deliver the cooling water at 20 liters per minute in this closed-cycle system.

(5) For question 4:

Yokoyama and Inoue repeatedly stress that their arc-melting for the $Zr_{55}Cu_{30}Ni_5Al_{10}$ alloy made an alloy ingot with good homogeneity, but they do not give any quantitative information. Such an assertion requires at least the element-mapping data obtained by reasonable means, e.g. FE-EPMA, SEM-EDX;WDX, WRF, etc., showing the homogeneous distribution of constituent elements in the resultant alloy buttons.

Yokoyama and Inoue[3] also stressed that the temperature gradient of the molten alloy ingot was estimated at about 110 K/mm and confirmed that the arc-melting method was suitable for the formation of the glassy phase. As discussed in the paper reported by Yokoyama et al.[5], the top surface temperature at the beginning of the cooling process is 1400 K for the $Zr_{50}Cu_{40}Al_{10}$ alloy button. Then, it is quite natural to assume that the cooling-start temperature of the $Zr_{55}Cu_{30}Ni_5Al_{10}$ alloy buttons is the same level, i.e., approximately 1400 K. Since Yokoyama and Inoue[3] report that the melting point of the $Zr_{55}Cu_{30}Ni_5Al_{10}$ alloy buttons is 1163 K, the liquid zone depth of the $Zr_{55}Cu_{30}Ni_5Al_{10}$ alloy buttons at the cooling-start time can be estimated by coupling the temperature gradient of the alloy melt. The liquid zone depth is estimated to be 2.15 mm from the simple relationship of (1400 K-1163 K)/110 K being shallow as expected. This value is found to be consistent with the cross-sectional photograph for the $Zr_{50}Cu_{40}Al_{10}$ alloy button, Fig.7 in the reference[5]. This seems to be a reason "why the alloy melt is viscous at the time of casting[3]". The top part (nearly half) of the arc-melted alloy buttons is considered in the two-phase state, i.e., solid phase plus liquid phase, because the bottom part, which is in direct contact with the copper hearth, is quite likely to be in the solid state. Such situation was previously described by Yokoyama et al.[6)] in terms of "pseudo float melting state".

Yokoyama and Inoue explain in their reply[3)] " *To ensure complete melting before casting, we carried out arc heating in two steps: (1) melting all the master alloy buttons together and (2) further melting of the molten alloy just at the pouring gate, as has been described in our earlier papers (Refs.7 and 8;cited in their report[2)])*". A need for the further melting at the pouring gate can be easily understood if the incomplete melting of alloy buttons is assumed in their experiments, whatever metal or alloy. In addition, the readers are not able to find any information for the use of an additional arc-rod to prevent hanging of molten alloy during the casting process in their earlier papers[7,8)] cited in their paper[2)] and reply[3)].

Casting of metallic melts under gravity is characterized by the kinematic viscosity, $\nu = \eta/\rho$, being in the range between 0.1 to $0.7 \times 10^{-6}$ m$^2$/sec, where $\eta$ and $\rho$ are viscosity and density of liquid metals (or alloys), respectively[7,8)] . This relationship is well-accepted for all metallic melts including alloys. Typical examples are suggested as follows. The kinematic viscosity is $1.0 \times 10^{-6}$ m$^2$/sec for drinking water, $H_2O$, at the ambient temperature and $0.7 \times 10^{-6}$ m$^2$/sec for liquid steel just above the melting temperature. For this reason, the kinematic viscosity of the $Zr_{55}Cu_{30}Ni_5Al_{10}$ alloy melt is assumed to be in the order of $1.0 \times 10^{-6}$ m$^2$/sec without any question. The readers' interest is to know the complete melting of the $Zr_{55}Cu_{30}Ni_5Al_{10}$ alloy ingot. From these comprehensive characteristics of liquid metals and alloys, there is no need for an additional heater at the pouring gate whenever the alloy ingot is in the perfect liquid state. In other words, the reply of Yokoyama and Inoue[3)] might give clear evidence for the imperfect melting of alloy ingots and the following imperfect casting of metallic glass rods.

(6) For question 5:

It is quite mysterious "why a new photograph of Fig.3[3)] was poorly prepared". The readers can certainly find defocusing, soft-focusing, and low contrast even relative to the previous patch-work photograph of Fig.4[2)] in 2007. In addition, the dark spot area of Fig.3 given in their reply [3)] does not match those shown in Fig.4 in the 2007 paper [2)].

Yokoyama and Inoue explain "*We did not have a stereoscope for observing the sample with a diameter of 30mm*". However, this is completely inconsistent with their other paper published in 2007 for the $Zr_{50}Cu_{40}Al_{10}$ alloy button[5)]. In this paper, Fig.7 shows the outer appearance of an arc-melted $Zr_{50}Cu_{40}Al_{10}$ alloy ingot about 32mm in diameter and its cross-sectional photograph (about 11 mm in height) is also provided. Note that the value of 32 mm is estimated from the scale of 5 mm given in the same photograph. This paper was published in 2007, earlier than the paper in question[2]. The readers can easily see that the area of Fig.7 in the reference[5], including a large sized alloy button, is about 46.3 x 46.3 $mm^2$. This fact is evidently inconsistent with their reply stating they did not have a large enough stereoscope for observing a sample with a diameter of 30mm.

(7) For question 6:

The catalog name of an HREM machine, JEOL-4000FX, also appears in another article[5]. If it is a typographical error, Yokoyama and Inoue need to correct all instances in their papers.

(8) For question 7:

The question is related only to the point that the HREM image of Fig.5[2] seems too smooth relative to the other HREM photographs of glassy alloys previously published. However, Yokoyama and Inoue refer only to the file size provided in the web-PDF files of Materials Transactions. Their reply is evidently inconsistent with the question. One possible reason may be the condition that the editorial committee of Japan Institute of Metals delivered the un-reviewed version of Kajitani's comments to Yokoyama and Inoue and thus unfortunately they may have prepared their reply to contents that were different from the published version[1]. This is quite likely to have happened as follows. The reply manuscript of Yokoyama and Inoue[3] was received on July 24, 2009, whereas the comment manuscript was accepted for publication on July 28, 2009. It is requested that the editorial committee find out the reason for this confusion.

The essential point of the question is again as follows. The readers can easily find that the HREM images of Zr-based metallic glasses using JEM-4000EX given in articles[2,5] are qualitatively very different, for example when the right half of Fig.13[5] is compared with Fig.5[2]. This difference is not understandable since HREM images of metallic glasses are known to be very similar. For this reason, the authors of the paper[2] are requested to give a clear answer about this point.

(9) For question 8:

The readers find it difficult to understand the statement given in the reply "*it (the suction-casting method) had several disadvantages (difficulty in de-molding, generation of* *cast defect etc.)*".

Inoue and Zhang[9] in 1966 definitely reported that production of bulk glassy $Zr_{55}Cu_{30}Ni_5Al_{10}$ alloy of 30 mm diameter by the suction casting method into a copper mold and no mention was made of cast defects etc. Yokoyama and Inoue as well as the other two authors should compare advantages, disadvantages, and limits for the sizes of metallic glassy rods between the suction casting method and the cap-casting one and also show "how and why the differences occurred". This is the normal method in research works. No reasonable answer is given in their reply[3] to the question "why is the suction casting method excluded in the conventional methods as mentioned in the paper[2]" and "why can they form the conclusion that the production of alloy glassy rods of 20mm in diameter is limited".

In conclusion, the reply[3] given by Yokoyama and Inoue contains many unclear points, so that the quality of the contents of their original paper[2] in 2007 is still a cause for concern.

REFERENCES
1) T.Kajitani: Mater. Trans. **50** (2009) 2502-2503.
2) Y.Yokoyama, E.Mund, A.Inoue and L.Schultz: Mater. Trans. **48** (2007) 3190-3192.
3) Y.Yokoyama and A.Inoue: Mater. Trans. **50** (2009) 2504-2506.
4) H.Kato, Y.Kawamura and A. Inoue: Mater. Trans. JIM **37** (1996) 70-77.
5) Y.Yokoyama, H.Fredriksson, H.Yasuda, M.Nishijima and A.Inoue: Mater. Trans. **48** (2007) 1363-1372.
6) Y. Yokoyama, K. Inoue and K. Fukaura: Mater. Trans. **43**(2002) 2316-2319.
7) Y. Kawai and Y. Shiraishi (Editors): Handbook of Physico-chemical Properties at High Temperature, Iron and Steel Institute of Japan, Tokyo, (1988) pp.1-23 and pp.93-20.
8) Japan Institute of Metals (Editor) : Handbook for Metals and Alloys, (4th edition), Maruzen, Tokyo, (2004) pp.63-69.
9) A. Inoue and T.Zhang: Mater. Trans. JIM **37** (1996) 185-187.